\documentclass[a4paper,11pt,titlepage,oneside,usenames,dvipsnames,svgnames,x11names,citecolor=NavyBlue,linkcolor=OliveGreen,urlcolor=blue]{article}
\pdfoutput=1
\usepackage{jheppub}
\usepackage{amsfonts, amsmath, amssymb}
\usepackage[normalem]{ulem}
\usepackage{quoting}
\usepackage{latexsym}
\usepackage{graphicx}
\usepackage{subcaption}
\usepackage[usenames,dvipsnames,svgnames,x11names]{xcolor}
\makeatletter
\def\@fpheader{\relax}
\makeatother

\def\al{\alpha}
\def\bt{\beta}
\def\gm{\gamma}

\def\TT{T\bar{T}}

\DeclareMathOperator{\csch}{csch}
\DeclareMathOperator{\arctanh}{arctanh}

\preprint{ CCTP-2024-5 \\ \hspace*{\fill}ITCP-IPP 2024/5}

\title{Bounds on $T\bar {T}$ deformation from  entanglement}
\author{Avik Banerjee$^{T}$,  and Pratik Roy$^{\bar{T}}$} 
\affiliation[T]{Crete Center for Theoretical Physics, Institute for Theoretical and Computational Physics,\\ Department of Physics, University of Crete, Heraklion, Greece} 
\affiliation[\text{\={T}}]{Mandelstam Institute for Theoretical Physics, School of Physics, NITheCS,\\ University of the Witwatersrand, Wits 2050, Johannesburg, South Africa} 
\emailAdd{avikphys02@gmail.com} 
\emailAdd{pratik.roy@wits.ac.za}

\abstract{
 Motivated by the existence of complex spectrum in $T\bar T$-deformed CFTs, in this paper we revisit the broadly studied topic of (holographic) entanglement entropy in the deformed theory to investigate its complex behaviour. As a concrete example, we show that in case of a 1+1 dimensional holographic CFT at finite temperature $\beta^{-1}$ and chemical potential $\Omega$, the holographic
 entanglement entropy in the deformed theory remains to be real only within the range  $-\frac{\beta^2}{8\pi^2}\frac{(1-\Omega^2)^2}{\Omega^2}< \mu < \frac{\beta^2}{8\pi^2}(1-\Omega^2) $ of the deformation parameter. While the upper bound overlaps with the familiar Hagedorn bound in the deformed theory, the novel lower bound on the negative values of the deformation parameter does not show up in thermodynamic quantities. However, from a holographic perspective we show that this intriguing lower bound is related to a spacelike to null transition of the associated Ryu-Takayanagi surface in the deformed geometry. We also investigate the Quantum Null Energy Condition in the deformed theory, within its regime of validity. }

\begin{document}
\maketitle

\section{Introduction}

Irrelevant deformations of local quantum field theories are in general poorly understood, in the sense that they typically destroy the existence of a UV fixed point and also spoil locality at scales set by the deformation parameter. In this context, the $T \bar T$ deformation of two-dimensional quantum field theories stands out as special. Firstly, this deformation is known to preserve integrability along the flow \cite{Smirnov:2016lqw}. Using integrability techniques, the S-matrix in the deformed theory can be computed exactly \cite{Dubovsky:2012wk} and it appears to be well-defined at all energies. Also, due to the remarkable factorization property of the expectation value of the deforming operator \cite{Zamolodchikov:2004ce}, the finite size spectrum in the deformed theory can be exactly computed as well. Given the exact knowledge of the finite size spectrum, the torus partition function in the deformed theory has also been studied \cite{Cardy:2018sdv} and it has been further shown that the torus partition function of a  $T\bar T$-deformed CFT is modular invariant despite the absence of conformal symmetry in the deformed theory \cite{Datta:2018thy}. Furthermore, it has been shown that in any two-dimensional quantum field theory with a dimensionful parameter $t$, demanding that the torus partition function is modular invariant and that its spectrum depends on the spectrum of $t=0$ theory in a universal way, uniquely fixes the finite $t$ theory to be a $T \bar T$ deformed CFT \cite{Aharony:2018bad}. 

Besides the remarkable field theoretic aspects, the $T \bar T$ deformation has gained a significant amount of attention in the context of string theory and holography.
 The deformed CFT has numerous stringy/gravity realizations,  in terms of non-critical string theory \cite{Callebaut:2019omt, Tolley:2019nmm}, CFT coupled to a 2D dilaton gravity theory \cite{Dubovsky:2017cnj, Dubovsky:2018bmo},  AdS$_3$  gravity with finite radial cut-off \cite{McGough:2016lol, Kraus:2018xrn, Taylor:2018xcy} (and its refinement \cite{Kraus:2022mnu}), and finally AdS$_3$  gravity with mixed boundary conditions \cite{Guica:2019nzm}. The last two realizations have rendered the deformation as a tractable irrelevant deformation of AdS$_3$/CFT$_2$ and triggered an extensive study of various aspects of the deformed theory using holographic techniques. For a pedagogical review of the history, developments, and modern aspects of the $T\bar T$ deformation, we refer to \cite{Jiang:2019epa,Guica:2022abc} and references therein.

One of the most investigated aspects of the deformed theory has been its entanglement properties \cite{Donnelly:2018bef,Chen:2018eqk,Gorbenko:2018oov,Murdia:2019fax,Ota:2019yfe,Banerjee:2019ewu,Jeong:2019ylz,He:2019vzf,Donnelly:2019pie,Asrat:2020uib,Allameh:2021moy,Setare:2022qls,He:2022xkh,Jeong:2022jmp,He:2023xnb}. In this paper, we are primarily interested in studying the holographic entanglement entropy (HEE) \cite{Ryu:2006bv,Hubeny:2007xt} in the deformed theory to investigate its reality as the deformation parameter is varied.  It is well known that given a finite deformation, the $T \bar T$-deformed CFT on a cylinder possesses a complex spectrum depending on the size of the cylinder and the scaling dimensions of the operators in the undeformed CFT. For such states, the von Neumann entropy will correspondingly be complex. In the discussions to follow, we shall consider the deformed theory to live on $\mathbb{R}^{1,1}$, where the spectrum is real. However, we will show that the HEE in a certain class of states can still become complex, and demanding its reality can put non-trivial bounds on the deformation parameter.

Building on the developments from the study of holographic entanglement entropy, \cite{Bousso:2015mna} proposed a generalization of the classical focusing theorem called the quantum focusing conjecture (QFC). The quantum null energy condition (QNEC), a non-gravitational limit of the QFC, states that the expectation value of the stress-energy tensor is bounded below by the second null variation of entanglement entropy. QNEC has been proven in various settings and to various degrees of generality \cite{Bousso:2015wca, Malik:2019dpg, Koeller:2015qmn, Balakrishnan:2017bjg, Ceyhan:2018zfg, Kudler-Flam:2023hkl}.\footnote{A generalization of QNEC \cite{Lashkari:2018nsl}, termed the Renyi QNEC, has also been proven to hold at least for free field theories \cite{Moosa:2020jwt,Roy:2022yzm}.} In \cite{Kibe:2021qjy}, non-violation of QNEC was shown to impose non-trivial constraints on the amount of growth of entropy and temperature upon a global homogeneous quench in a thermal two-dimensional CFT. \cite{Banerjee:2022dgv} extended this study to global inhomogeneous quenches, and using QNEC found evidence that two-dimensional holographic CFTs can be used to create erasure tolerant quantum memory. Motivated by these studies, we investigate if  QNEC remains valid in the deformed theory as well and, in particular, whether its non-violation imposes non-trivial constraints on the flow.

However, a potential issue in the study of QNEC in the deformed theory is related to the non-locality. While QNEC is valid for local QFTs, the deformed theory is known to be non-local at high energies, with the scale of non-locality being set by the deformation parameter to be $\sqrt \mu $. To circumvent this, we shall restrict our discussion of QNEC to scales much greater than $\sqrt \mu$, so that the theory can be treated quasi-locally. In particular, we shall assume that there exists a stress-tensor in the deformed theory and that the entanglement entropy of a region of length $l \gg \sqrt{\mu}$ is well-defined. Since these quantities have been rigorously studied, we believe that these underlying assumptions for studying QNEC in the deformed theory are mild and justifiable.

\newpage
Our main results are as follows. Working within the holographic mixed boundary condition interpretation of the deformation, we show that in a $T\bar T$-deformed  1+1 dimensional holographic
CFT at temperature $T^{-1}=\beta$ and chemical potential $\Omega$ due to a boost, the reality of holographic entanglement entropy puts sharp bounds on the deformation parameter for both signs, confining it to lie within the range 
\begin{align} \nonumber
    -\frac{\beta^2(1-\Omega^2)^2}{8\pi^2 \Omega^2} < \mu < \frac{\beta^2(1-\Omega^2)}{8\pi^2}.
\end{align}
In the case of vanishing $\Omega$, the lower bound disappears while the upper bound coincides with the familiar Hagedorn bound in the deformed theory. The non-trivial lower bound in the presence of a finite chemical potential is typically not perceived by the thermodynamic quantities.  However, holographic computation reveals that this lower bound follows from a spacelike to null transition of the Ryu-Takayanagi surface in the deformed geometry. Within this allowed regime, we find no violations of QNEC in the deformed theory.

The rest of the paper is organized as follows. In Section \ref{Sec2}, we briefly review the variational principle approach to derive the deformed metric and the stress-tensor one-point function along the flow, which eventually leads to the two holographic descriptions of the deformation. In Section \ref{Sec3}, we compute the HEE in the deformed theory, both in the absence and presence of a chemical potential associated with a boost, and derive our bounds on $\mu$ based on the reality of HEE. In Section \ref{Sec4}, we study the QNEC inequality in the deformed theory. Finally, we end with a discussion in Section \ref{Sec5}. A couple of appendices are also provided for computational details.

\section{$T\bar{T}$ deformation and  its holographic descriptions }\label{Sec2}

In this section, we briefly review the variational principle approach of \cite{Guica:2019nzm} to obtain the deformed metric and the stress-tensor one-point function in the $T\bar{T}$-deformed two-dimensional conformal field theories. For a detailed discussion on this approach, we refer to \cite{Bzowski:2018pcy}, and to \cite{Papadimitriou:2007sj} for a more general discussion. 

The $T\bar{T}$ deformation is a universal irrelevant deformation of any local two-dimensional (conformal) quantum field theory, defined by the differential relation
\begin{equation}
    \partial_{\mu}S^{[\mu]}= -\frac{1}{2} \int~ d^2x \sqrt{\gamma} \left(T^{\alpha\beta}T_{\alpha\beta}-T^2\right)_{\mu}, \label{defttbar}
\end{equation}
where the quantities appearing on the right-hand side correspond to the deformed theory and we are working in the Euclidean signature following \cite{Guica:2019nzm}. The deformation generates an integrable flow in the space of field theories where at any arbitrary point, the stress tensor $T_{\alpha\beta}^{[\mu]}$ (and hence the deforming operator) is to be computed from the response of $S^{[\mu]}$ to an arbitrary variation in the metric. This assertion in turn restricts our discussions to scales much greater than $\sqrt \mu$, where the theory can be treated as quasi-local.

Given the fact that $T\bar{T}$  is a double-trace deformation where the deforming operator also depends on the source, it is natural to expect the holographic dictionary for the deformed theory to be modified. In fact, such deformations are known to lead to mixed boundary conditions \cite{Witten:2001ua}. In the context of $T\bar{T}$ deformation, the deformed dictionary can be most efficiently worked out, at least for large $N$ CFTs, using the variational principle approach \cite{Guica:2019nzm}. 
Taking a variation of the defining relation \eqref{defttbar} with respect to the metric and after some simple algebra, one arrives at
\begin{align} \label{variation}
 \partial_{\mu}\left(\sqrt{\gamma}  T_{\al\bt}\right)&\, \delta \gamma^{\al\bt} + \sqrt{\gamma} T_{\al\bt} \partial_{\mu} \left(\delta \gamma^{\al\bt}\right) \\
	 =&\ \sqrt{\gamma}\left[\left(-\frac{1}{2} \gamma_{\al\bt} \mathcal{O}_{T\bar{T}}-2 T_{\al\gamma} T_{\bt}^{\gm} +2 T T_{\al\bt}\right) \delta \gamma^{\alpha \beta}+ 2 T_{\al\bt} \delta \left(T^{\al\bt} - \gamma^{\al\bt} T\right)\right],   \nonumber
\end{align}
where we have used $$\delta S^{[\mu]}=
\left(\frac{1}{2} \int d^2 x~ \sqrt{\gamma}~T_{\alpha\beta}\delta\gamma^{\alpha\beta}\right)^{[\mu]},$$ and $T_{\al\bt}$ stands for its expectation value for the rest of the paper.
Comparing the quantity being varied and the coefficient of the variation in \eqref{variation}, one arrives at the flow equations
\begin{equation} \label{fe2}
\partial_{\mu} \gm^{\al\bt} = 2 \left(T^{\al\bt} -\gm^{\al\bt} T\right), \qquad \partial_{\mu}\left(\sqrt{\gamma} T_{\al\bt}\right) =\sqrt{\gm}( 2 T T_{\al\bt} - 2 T _{\al\gm}T_{\bt}^{\gm} -\frac{1}{2}\gamma_{\al\bt}\mathcal{O}_{T\bar{T}}),
\end{equation}
where $\mathcal{O}_{TT}= T^{\alpha\beta}T_{\alpha\beta}-T^2$. Using the notation $\hat{T}_{\al\bt}= T_{\al\bt}- \gm_{\al\bt}T$, these equations can be written in a more compact form as
\begin{align}
\partial_{\mu} \gm_{\al\bt} =&\ -  2 \hat{T}_{\al\bt},  \label{fe21} \\
\partial_{\mu} \hat{T}_{\al\bt} =&\  -\hat{T}_{\al\gm} \hat{T}_{\bt}^{\gm}, \label{fe22}
\end{align}
Note that we have used the relations $\partial_{\mu}\sqrt{\gm} = \sqrt{\gm} T$, $T T_{\al\bt} -T_{\al\gm}T_{\bt}^{\gm}= -\frac{1}{2} \gm_{\al\bt} \mathcal{O}_{T\bar{T}}$, and $\partial_{\mu} T = -T^{\al\bt} T_{\al\bt}$ to get to (\ref{fe22}). Differentiating (\ref{fe21}) twice and using (\ref{fe22}), we arrive at the final form of the flow equations 
\begin{align} 
\partial_{\mu}^3 \gm_{\al\bt} =&\  0, \label{fe31} \\
\partial_{\mu} \hat{T}_{\al\bt} =&\  -\hat{T}_{\al\gm} \hat{T}_{\bt}^{\gm},
\label{fe32}
\end{align}
where to arrive at \eqref{fe31}, we have used\footnote{See  Appendix \ref{Id} for brief proofs of various identities.} \begin{equation}
    \partial_{\mu}( \hat{T}_{\al\gm} \hat{T}_{\bt}^{\gm})  = 0.
\end{equation}
Since $\hat{T}_{\al\gm} \hat{T}_{\bt}^{\gm}$ is $\mu$ independent, the flow equations \eqref{fe31}-\eqref{fe32}  can now be easily solved to get the deformed metric and stress-tensor expectation value as
\begin{align} \gm^{[\mu]}_{\al\bt} =&\  \gm^{[0]}_{\al\bt} - 2 \mu  \hat{T}^{[0]}_{\al\bt} +\mu^2  \hat{T}^{[0]}_{\al\rho} \gamma^{[0]\rho\sigma}\hat{T}_{\sigma \bt}^{[0]}\, , \label{solfe1}\\
 \hat{T}^{[\mu]}_{\al\bt} =&\  \hat{T}^{[0]}_{\al\bt} -\mu \hat{T}^{[0]}_{\al\rho} \gamma^{[0]\rho\sigma}\hat{T}_{\sigma \bt}^{[0]}\,,  \label{solfe2}
\end{align}
where the quantities with superscript $[0]$ refer to the undeformed theory. Note that these solutions are exact in the deformation parameter, not perturbative. Let us now plug back the undeformed dictionary in \eqref{solfe1}. In the case of pure gravity, identifying
\begin{equation}
    \gamma^{[0]}_{\alpha\beta} \equiv g^{(0)}_{\alpha \beta}\,,\qquad \hat{T}^{[0]}_{\alpha \beta}\equiv  \frac{1}{8\pi G l_{AdS}}  g^{(2)}_{\alpha \beta}\,, \qquad \hat{T}^{[0]}_{\al\rho} \gamma^{[0]\rho\sigma}\hat{T}_{\sigma \bt}^{[0]} \equiv \frac{1}{(4\pi G l_{AdS})^2} g^{(4)}_{\alpha \beta}\, ,  \label{undefmap}
\end{equation}
as coefficients of the Fefferman-Graham expansion for the solution of Einstein's equations with $\Lambda = -2/l_{AdS}^2$,
\[
    ds^2 = l_{AdS}^2 \frac{d\rho^2}{4 \rho^2}+ \left(\frac{g^{(0)}_{\alpha \beta}}{\rho}+g^{(0)}_{\alpha \beta}+ \rho~ g^{(4)}_{\alpha \beta}\right) dx^{\alpha} dx^{\beta},
\] 
we readily note that fixing the deformed metric in \eqref{solfe1} now amounts to imposing Dirichlet boundary condition at a finite radial slice $\rho_c= -\frac{\mu}{4\pi G l_{AdS}}$. In this case, one can further show \cite{Guica:2019nzm} that the deformed stress-tensor\footnote{The physical stress-tensor is $T^{[\mu]}_{\al\bt}$, not $\hat{T}^{[\mu]}_{\al\bt}$.}  now has the identification 
\begin{equation*}
    {T}^{[\mu]}_{\al\bt}= T^{BY}_{\al\bt}(\rho_c)-\frac{g_{\al\bt}(\rho_c)}{8\pi Gl_{AdS}}.
\end{equation*}
where $ T^{BY}_{\al\bt}(\rho_c)$  and $g_{\al\bt}(\rho_c)$ are the Brown-York stress-tensor and the induced metric evaluated at $\rho=\rho_c$, respectively. The second term ensures that when $\rho_c \rightarrow 0$ (and hence $\mu \rightarrow 0 $), the CFT stress-tensor precisely agrees with the Krauss-Balasubramanian proposal \cite{Balasubramanian:1999re}. 

However, these identifications of both the deformed metric and stress-tensor expectation value are true only for the negative sign of the deformation parameter and in the case of pure gravity. In the presence of matter, the last of the identifications made in \eqref{undefmap} does not hold, rendering the finite cut-off prescription ambiguous. Nevertheless, just from the first two identifications in \eqref{undefmap}, which still remain valid in the presence of matter, we can infer that fixing $\gm^{[\mu]}$ can alternatively be thought as a mixed boundary condition involving both $g^{(0)}$ and $g^{(2)}$  at the original boundary as compared to a Dirichlet boundary condition on $g^{(0)}$ in the undeformed case.

\subsection{The deformed spacetime}

 In what follows, we shall compute holographic entanglement entropy and evaluate the QNEC inequality in the deformed theory primarily within the mixed boundary condition (MBC hereafter) interpretation, which gives access to both signs of the deformation. Since the computations will be holographic, in this section, we work out the relevant deformed spacetime subject to a fixed deformed boundary metric\footnote{In what follows, we shall assume the deformed theory to live on $\mathbb{R}^{1,1}$.},  say $\gm^{[\mu]}_{\al\bt}=\eta_{\al\bt}$, in some null coordinates $(U, V)$. Since $\sqrt{-\gm}R$ is invariant along the flow \cite{Guica:2019nzm}, for a Ricci flat deformed metric, the undeformed metric must also be Ricci flat. We can again choose this Ricci flat undeformed metric to be $\eta_{\al\bt}$ in some different null coordinates, say $(u,v)$. For the undeformed case, for a flat boundary metric $g^{[0]}_{\al\bt}=\eta_{\al\bt}$, the most general solution of Einstein's equations with a negative cosmological constant $\Lambda= -2/l_{AdS}^2$ is given by the Banados spacetime,
 \begin{equation}
     ds^2 = l_{AdS}^2 \frac{d\rho^2}{4 \rho^2}+\frac{du dv}{\rho}+\left(L(u) du^2 +\bar L (v) dv^2\right)+ \rho  L (u) \bar L (v)~ du dv, \label{undefbanados}
 \end{equation}
 which is characterized by two chiral functions $L(u)$ and $\bar L (v)$. The induced metric at a slice $\rho=\rho_c$ is simply given by
 \begin{equation}
    \rho_c ds^2_{\rho_c}= \left(du + \rho_c \bar{L}(v) dv\right)\left(dv + \rho_c {L}(u) du\right). \label{metricfc}
 \end{equation}
 In the finite cut-off picture, the deformed metric lives on this $\rho=\rho_c$ hypersurface. Note that the induced metric \eqref{metricfc} can be cast into the desired (deformed) flat metric $ds^2_{[\mu]}=dU dV$ with the help of the state-dependent coordinate transformations \cite{Guica:2019nzm}
 \begin{equation} \label{pcfmap1}
     U= u + \rho_c \int^v\bar{L}(v') dv'~~,~~~ V= v + \rho_c\int^u L(u') du'.
 \end{equation}
 In general, this map is not invertible. However, in the discussions to follow, we shall consider the chiral functions to be constants, $L(u)=L$ and $\bar L(v)= \bar L$, allowing us to invert the map to get
 \begin{equation}
     u= \frac{U -  \rho_c \bar{L} ~V}{1-\rho_c^2 L \bar{L}}\ ,\qquad v= \frac{V - \rho_c L~ U}{1-\rho_c^2L \bar{L}}~. ~~ \label{pcfmap2}
 \end{equation}
 Then the Fefferman-Graham expansion for the spacetime with a flat metric at $\rho=\rho_c$ is simply obtained by using the map \eqref{pcfmap2} in the Fefferman-Graham expansion of the undeformed spacetime \eqref{undefbanados}, giving
\begin{equation}		\label{defbanados1}
ds^2_{def} = l_{AdS}^2\frac{d\rho^2}{4\rho^2}+\frac{ds^{(0)2}_{def}}{\rho} + ds^{(2)2}_{def}+ \rho ~ds^{(4)2}_{def}~, 
\end{equation}
where,
\begin{align}   
ds^{(0)2}_{def}=&\   \frac{\left(dU - \rho_c \bar L ~dV\right)\left(dV - \rho_c {L} ~dU\right)}{(1-\rho_c^2 L \bar{L})^2} ,  \nonumber\\
ds^{(2)2}_{def}= &\ \frac{\left(1+ \rho_c^2 L \bar{L}\right)\left(L ~dU^2+ \bar{L}~dV^2\right)-4\rho_c L \bar L~dUdV}{(1-\rho_c^2 L \bar{L})^2}, \nonumber\\
ds^{(4)2}_{def}=&\  L \bar{L} ds^{(0)2}_{def}.	\label{defbanados2}
\end{align}
with $\rho \geq \rho_c$. To obtain the relevant spacetime for the MBC interpretation, all we need to do is make the identification 
\begin{equation}
    \rho_c =-\frac{\mu}{4\pi Gl_{AdS}} = -2\mu
\end{equation}
in \eqref{defbanados2} and then allow for both signs of $\mu$. Note that here and in the rest of the paper, we set $8\pi G =1$ and the AdS radius $l_{AdS}=1$.\footnote {These two choices also fix $c=12\pi$.} The deformed spacetime is again given by \eqref{defbanados2}, but this time with $\rho \geq 0$ and  the coefficients of FG expansion explicitly given by 
\begin{align}
ds^{(0)2}_{def}=&\   \frac{\left(dU + 2\mu  \bar L~ dV\right)\left(dV + 2\mu {L} ~dU\right)}{(1-4 \mu^2 L \bar{L})^2} , 
\nonumber\\
ds^{(2)2}_{def}= &\ \frac{\left(1+ 4\mu^2 L \bar{L}\right)\left(L dU^2+ \bar{L}dV^2\right)+8\mu  L \bar L~dUdV}{(1-4 \mu^2 L \bar{L})^2}, \nonumber\\
ds^{(4)2}_{def}=&\  L \bar{L} ds^{(0)2}_{def}.	\label{defbanados3}
\end{align}
Note that in \eqref{defbanados3} and in the discussions to follow, $L$ and $\bar L$ will stand for the deformed quantities $L^{\mu}$ and $\bar L^{\mu}$ respectively.  One can now easily check that
$$ds^{(0)2}_{[\mu]}- 2\mu ds^{(2)2}_{[\mu]}+ 4\mu^2 ds^{(4)2}_{[\mu]}=dUdV.$$
Finally, we conclude the section by reporting the expectation value of the deformed stress tensor which will be relevant for the QNEC computation. In  $(U,V)$ coordinates, the components are given by\footnote{Recall we have set $8\pi G=1$ and $l_{AdS}=1$.} 
\begin{align} \label{stress-tensor}
    T^{[\mu]}_{\al\bt} =\frac{1}{\left(1-4\mu^2 L \bar{L}\right)}\begin{pmatrix}
L & -2\mu L \bar{L}\\
-2\mu L \bar{L} & \bar{L}\\
    \end{pmatrix}.
\end{align}
 Note that the deformed stress tensor now has an off-diagonal component, which is however not independent.

\section {Holographic Entanglement Entropy in the deformed theory} \label{Sec3}

In this section, we will work out the HEE of a region of length $l$ at finite temperature $T=\beta^{-1}$ in the deformed theory, both in the absence and presence of a chemical potential $\Omega$ associated with a boost. In both cases, we shall use the MBC interpretation for the holographic description and compute geodesic lengths in the relevant deformed spacetimes. 

\subsection{Vanishing chemical potential}

To illustrate the effect of the chemical potential, let us begin with the case when it is absent. In this case, the dual spacetime is the deformed non-rotating planar BTZ, and from (\ref{defbanados1}) and (\ref{defbanados2}) with $\bar L=L$,  it is given by
\begin{equation}
ds^2= \frac{d\rho^2}{4 \rho^2} 	 + \frac{L(\rho+ 2 \mu)(1 +2 \mu L^2 ~\rho)}{\rho (1-4 \mu^2 L^2)^2}\left(dU^2+dV^2\right)+\frac{1+4  \mu^2 L^2+ L^2 \rho\left(\rho+ 8\mu +4 \mu^2 L^2 \rho \right)}{\rho (1-4 \mu^2 L^2)^2}\, dUdV.
\end{equation}
 With the transformations
\begin{equation}
\nonumber \rho=\frac {r^2 - 2L- r \sqrt{r^2-4L}}{2L^2},\quad dU=dX+dT,\quad dV=dX-dT,
\end{equation}
this metric can be recast into the standard non-rotating BTZ form,
\begin{equation}	\label{defbtz}
ds^2= \frac{dr^2}{r^2-4L} - (r^2-4L)~\frac{ dT^2}{\left(1+2\mu L\right)^2}  + r^2~\frac{dX^2}{ \left(1-2\mu L\right)^2},
\end{equation}
which has a horizon at $r_h= 2\sqrt L$. Note that this horizon is now $\mu$ dependent. Thus the coordinate transformations \eqref{pcfmap2} have the effect of rescaling both time and space, with different scale factors. Here we emphasize that the  Fefferman-Graham radial coordinate $\rho$ ranges between $0\leq\rho \leq 1/{L},$ and consequently the BTZ radial coordinate $r$  has the range $\infty > r \geq r_h$.

To compute HEE, we use the method developed in \cite{Kibe:2021qjy}. Consider an interval on the boundary of (\ref{defbtz}) delimited by the points $p_1=(r_c,0,0)$ and $p_2=(r_c,0,l)$, where $r_c \rightarrow \infty$. In order to compute HEE in the deformed geometry (\ref{defbtz}), we perform a series of coordinate transformations (detailed in Appendix \ref{maps}) that maps \eqref{defbtz}  to the Poincare patch of AdS$_3$,
 \begin{equation} \label{Poincare}
ds^2=\frac{-2 dZ dW -dW^2 +dY^2}{Z^2}~.
\end{equation}
In the Poincare patch, a closed-form expression is known for the length of geodesics between two arbitrary points. Let $P_1=(Z_1,W_1,Y_1)$ and $P_2=(Z_2,W_2,Y_2)$ be the final image of $p_1$ and $p_2$ respectively in (\ref{Poincare}). The geodesic length connecting these two points is then given by
\begin{equation} \label{length}
    \mathcal{L}=\log\left(\zeta+\sqrt{\zeta^2 -1}\right),
\end{equation}
where $\zeta$ is the unique conformal invariant associated with any two points in the Poincare patch of AdS$_3$ \cite{Polchinski:2010hw}. This invariant is given by
\begin{equation} \label{invariant}
\zeta= \frac{1}{2Z_1Z_2}\left(Z_1^2+Z_2^2-\left(Z_1+W_1-Z_2-W_2\right)^2+\left(Y_1-Y_2\right)^2\right).
\end{equation}
For spacelike geodesics, $\zeta>1$, whereas $\zeta=1$ corresponds to null geodesics. Using  the maps (\ref{defbtzrescale})-(\ref{EtoP}), it can be calculated to be
\begin{equation}\label{zeta}
    \zeta= \frac{2 \sinh^2{\frac{l\sqrt L}{1-2\mu L}}+4z_c^2L}{4z_c^2L},
\end{equation}
with $z_c=1/r_c$. Now in the  MBC picture, the geodesic endpoints are at $z_c \rightarrow 0$. In this limit, $\zeta 
 \sim \mathcal{O}(1/z_c^2)$ from \eqref{zeta}, and we can approximate $\sqrt{\zeta^2-1}$ in \eqref{length} by $\zeta$. Then, from \eqref{length} the geodesic length is given by  
\begin{equation}\label{Lnonrot}
\mathcal{L}= \log \left(2\frac{  \tilde{\zeta}}{z_c^2}\right), \qquad \tilde{\zeta}= \frac{2\sinh^2{\frac{l\sqrt L}{1-2\mu L}}}{4 L}.
\end{equation}
Note that in terms of $\tilde{\zeta}$, spacelike geodesics will correspond to $\tilde {\zeta} > z_c^2/2$, whereas $\tilde{\zeta} \rightarrow z_c^2/2$ is the null limit, with $z_c \rightarrow 0$. So to a good approximation, we will consider $\tilde {\zeta} \to 0 $ as the null limit.

In the finite cut-off picture, neither of these approximations would have been valid with $z_c$ being finite\footnote{$z_c$ is related to the Fefferman-Graham radial cut-off $\rho_c$ as $z_c=\frac{\sqrt{\rho_c}}{1+\rho_cL}$.}. In that case, one needs to use \eqref{length} along with \eqref{zeta} exactly to compute the geodesic length. It is then natural to expect that the qualitative features of the resulting geodesic lengths and therefore of the HEE will be quite different. For example, for finite $z_c$, if we take the limit $l\rightarrow 0$, then  from \eqref{zeta} $\zeta \rightarrow 1$ and hence $\mathcal{L} \rightarrow 0$. However, in the MBC picture, the $l \rightarrow 0$  limit of $\mathcal{L}$ is $z_c$ dependent.

Finally, the HEE in the deformed theory is given by
\begin{equation}\label{entbtz}
S_{\text{ent}}=\frac{c}{6}  \mathcal{L} = \frac{c}{3} \log\left(\frac{\left(\beta + \sqrt{\beta^2 -8\pi^2 \mu }\right)\sinh \left(\frac{\pi l}{\sqrt{\beta^2 -8\pi^2 \mu}}\right)}{2 \pi z_c}\right),
\end{equation}
where we have used
\begin{align} \label{defL}
L=\frac{\beta^2 -4\pi^2 \mu-\beta \sqrt{\beta^2 -8\pi^2 \mu}}{8\pi^2\mu^2}.
\end{align}
The last relation follows from inverting $\beta = \frac{\pi}{\sqrt{L}}\left(1+2 \mu L\right)$, which in turn follows from computing the Hawking temperature from \eqref{defbtz} and identifying the same with the temperature $\beta$ of the undeformed theory. Note that the HEE computed by this method agrees exactly with the non-perturbative result obtained in \cite{He:2023xnb} using the gravitational Wilson line technique, with $\mu$ replaced by $-\mu/2$. In particular, the reality of HEE puts a sharp upper-bound\footnote{Due to the relative sign between the deformation parameter used in this paper and the one used in \cite{He:2023xnb}, our upper bound on the positive values of the deformation parameter \eqref{ubound} translates to a lower bound on the negative values of the deformation parameter in \cite{He:2023xnb}.}
\begin{equation} \label{ubound}
    \mu <  \mu_{u}= \frac{\beta^2}{8\pi^2}.
\end{equation}
Note that this upper bound is analogous to what follows from the reality of the ground state energy in the deformed CFT on a cylinder, with $\beta$ replaced by the size of the cylinder $R$ and $c=12\pi$. In fact, this upper bound is simply the Hagedorn bound written in terms of $\mu$. Thermal equilibrium at temperature $T=\beta^{-1}$ ceases to make sense for $\mu >\mu_u$. 

\begin{figure}[tp]
    \centering
\includegraphics[scale=1]{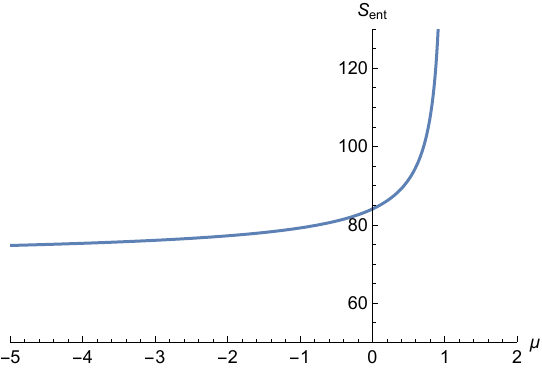}
    \caption{{\small Plot for HEE as a function of the deformation parameter $\mu$. Here $l=5,~ \beta = \sqrt{8}\pi, z_c=0.01$. HEE becomes complex beyond $\mu =\mu_u (=1)$, but continues to be real for arbitrary large negative $\mu$.} }
    \label{Fig:1}
\end{figure}

Note that for $\mu <\mu_u$, the argument of the logarithm in \eqref{entbtz} is always greater than zero. Since this argument is proportional to $\tilde{\zeta}$, in this case, the geodesic always remains spacelike. Figure \ref{Fig:1} shows the variation of HEE with the deformation parameter $\mu$ for fixed $l$ and $\beta$. Clearly, HEE continues to be real for arbitrarily large negative values of $\mu$. However, in the next subsection, we will see that the presence of a chemical potential puts a sharp lower bound on negative values of $\mu$ as well.

Finally, for small $\mu$, we have 
\begin{align} \label{entbtzser}
S_{\text{ent}}=& \frac{c}{3} \log \left(\frac{\beta  \sinh \left({\pi  l}/{\beta }\right)}{ \pi z_c}\right)+ \frac{2 c \pi ^2    }{ 3 \beta ^2} \left(2 \left(\frac{\pi  l}{\beta }\right)\coth \left(\frac{\pi  l}{\beta }\right)-1\right) \mu +\mathcal{O}(\mu^2).
\end{align}
The first term gives the entanglement entropy in the undeformed CFT if we identify $\epsilon= z_c$ as the UV cut-off. The second term gives the leading correction due to the $\TT$ deformation. This correction term precisely agrees\footnote{To check this, one must replace $\mu \rightarrow -\frac{\mu \pi c}{12}$ in \eqref{entbtzser}  because of the difference in conventions.} with the replica trick result \cite{Chen:2018eqk} in the high-temperature limit $\beta \ll l$.

Before moving on to the next section, it is worth comparing the results to the ones obtained in the finite cut-off prescription. In the finite cut-off approach, one should use \eqref{length} exactly, with $\zeta$ given by \eqref{zeta}. As expected, the resulting expression for HEE does not agree with \eqref{entbtz}. However, in the small $\rho_c$ limit, we have
\begin{align}
    S_{\text{ent}}^{\text{FC}} = \frac{c}{3} \log \left[\frac{\beta  \sinh \left({\pi  l}/{\beta }\right)}{ \pi \sqrt{\rho_c}}\right]- \frac{ c \pi ^2   }{ 3 \beta ^2} \left( 2 \left(\frac{\pi  l}{\beta }\right)\coth \left(\frac{\pi  l}{\beta }\right)-\left(1+\coth^2\left(\frac{\pi l}{\beta}\right)\right)\right) \rho_c +\mathcal{O}(\rho_c^2).
\end{align}
The leading term agrees with its counterpart in the MBC prescription upon identifying $\sqrt{\rho_c}=\epsilon$ as the UV cutoff. However, there is a clear mismatch in the subleading term after identifying $\rho_c=-2\mu$. So, even in the absence of matter and for negative values of the deformation parameter, the two prescriptions do not agree when one considers non-local quantities like entanglement entropy. However, if we consider the high-temperature limit $l \gg \beta$, then the subleading term agrees precisely.

\subsection{Finite chemical potential}

\begin{figure}[!h]
    \centering
    \includegraphics[width=.65\textwidth]{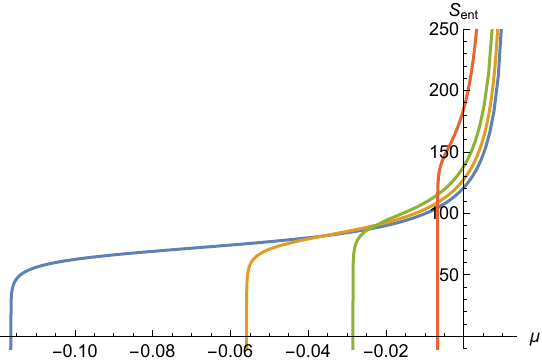}
    \caption{{\small Plot for HEE as a function of the deformation parameter $\mu$ for various values of $\Omega$. The blue curve corresponds to $\Omega=0.3$, orange curve corresponds to $\Omega=0.4$, green curve corresponds to $\Omega =0.5$, and the red curve corresponds to $\Omega=0.7$. For all the plots, we have used   $l=2,~ \beta = 1, z_c=0.01$. The lower bound on negative $\mu$ becomes monotonically stricter as $\Omega$ increases. }}
    \label{Fig:2}
\end{figure}

Next we consider the effect of a finite chemical potential by boosting the thermal state. In this case, the dual spacetime is a deformed boosted (planar) BTZ black hole with $L \neq \bar{L}$. In Fefferman-Graham gauge, the deformed spacetime takes the following form,
\begin{align}
    ds^2 =&\ \frac{d\rho^2}{4\rho^2} +\frac{1+4 \mu^2 L \bar L+ L \bar L \rho \left(\rho+8\mu+ 4 \mu^2 L \bar L~ \rho\right)}{\rho(1-4 \mu^2 L \bar L)^2}\, dUdV  +\frac{(\rho+2 \mu)(1+ 2\mu L \bar L ~\rho)}{\rho (1-4 \mu^2 L \bar L)^2}\left(L dU^2+ \bar L dV^2\right). 
\end{align}
Writing
\begin{equation}
    L=\frac{1}{4}(r_+- r_-)^2, \qquad \bar L=\frac{1}{4}(r_++ r_-)^2, \qquad r_h^2=r_+^2-r_-^2,
\end{equation}
and defining new coordinates
\begin{align}
    \rho=\ \frac{r^2-(L+\bar L)-\sqrt{\left(r^2-(L+\bar L)\right)^2-4 L\bar L}}{2 L \bar L} ,
    \quad U= X+T,\quad V= X-T,
\end{align}
this metric can be recast into the following  form,
\begin{equation} \label{defrotbtz}
    ds^2 = \frac{r^2dr^2}{(r^2-r_+^2)(r^2-r_-^2)} - \frac{4(r^2-r_+^2)(r_+dT - r_-dX)^2}{r_h^2(2-r_h^2\mu)^2} + \frac{4(r^2-r_-^2)(r_-dT-r_+dX)^2}{r_h^2(2+r_h^2\mu)^2}  .
\end{equation}
This geometry is dual to a $T\bar{T}$-deformed 1+1 dimensional CFT at a finite temperature $\beta$ and chemical potential $\Omega$ given by
\begin{align} \label{betaomega}
    \beta = \frac{\pi}{2} \left(\frac{1}{\sqrt L}+ \frac{1}{\sqrt {\bar L}}\right)\left(1+2 \mu \sqrt{L \bar L}\right), \qquad \Omega= \frac{\sqrt {\bar L}-\sqrt L}{\sqrt {\bar L}+\sqrt L}.
\end{align}
To compute HEE in this geometry, we again choose a region on the boundary of (\ref{defrotbtz}) delimited by the points $p_1=(r_c,0,0)$ and $p_2=(r_c,0,l)$. To use the working formulae (\ref{length})-(\ref{invariant}), we again need to put (\ref{defrotbtz}) into the form (\ref{Poincare}) as detailed in  Appendix \ref{maps}. Then computing the invariant, $\zeta,$ and ignoring terms as described in the previous subsection, we arrive at the following non-perturbative expression for HEE,
\begin{equation} \label{entrotbtz}
    S_{\text{ent}}= \frac{c}{6}\log\left(8\pi^2\mu^2 \frac{\cosh{\left(\frac{\pi l \left(\beta_++\beta_-\right)}{\beta_+\beta_-\sqrt{1- \frac{8\pi^2 \mu}{\beta_+\beta_-}}}\right)}-\cosh{\left(\frac{\pi l\left(\beta_+-\beta_-\right)}{\beta_+\beta_-}\right)}}{z_c^2 \beta_+ \beta_-\left(1-\sqrt{1-\frac{8 \pi^2 \mu}{\beta_+\beta_-}}\right)^2}\right),
\end{equation}
where $\beta_{\pm}= \beta(1 \pm \Omega)$. 
Let us now analyze the reality conditions of $S_{\rm ent}$.  Firstly, the upper bound which follows from the vanishing of the expression appearing under the square roots  now gets modified to 
\begin{align} \label{uboundrotbtz}
    \mu_{u} = \frac{\beta_+\beta_-}{8\pi^2}.
\end{align}
Now for $\mu <\mu_u$, when $\beta_-\neq \beta_+$, the argument of the logarithm has a  zero when the two terms in the numerator become equal. This happens precisely when $\mu$ equals 
\begin{equation} \label{lboundrotbtz} 
\mu_{l} =-\frac{\beta_+^2 \beta_-^2}{2\pi^2 (\beta_+-\beta_-)^2 }. 
\end{equation}
This is the value of $\mu$ at which the spacelike RT surface connecting the two endpoints becomes null. We emphasize here that in the case of a vanishing $\mu$, the numerator is always greater than zero for finite  $\beta_+ \neq \beta_-$ (finite $\Omega$). Even in the presence of a finite $\mu < \mu_u$, the numerators would have always been greater than zero for $\beta_+=\beta_- > 0$. So the spacelike to null transition happens only in the presence of a finite $\mu$ along with $\beta_+ \neq \beta_-$. Finally, expressing both the bounds in terms of $\beta$ and $\Omega$, the reality of HEE confines the allowed values of the deformation to be within the window
\begin{align} \label{murange}
    -\frac{\beta^2(1-\Omega^2)^2}{8\pi^2 \Omega^2} < \mu < \frac{\beta^2(1-\Omega^2)}{8\pi^2}.
\end{align}
In the limit $\Omega \rightarrow 0$, the upper bound reduces to \eqref{ubound} while the lower bound becomes trivial. Figure \ref{Fig:2} shows the plot for HEE against $\mu$ for various values of $\Omega$. Note that,  $\mu_l$ corresponds to the value of $\mu$ where $S_{\text{ent}}$ diverges to $- \infty$, eventually becoming complex thereafter. However, before that, there exists a small window where the geodesic length remains negative, which is also unphysical. So one can alternately propose a lower bound $\mu_l^*$ corresponding to the vanishing of the HEE. However, in the limit $z_c \rightarrow 0$, the difference between $\mu_l$ and $\mu_l^*$ becomes vanishingly small. This is illustrated in Figure \ref{Fig:2}, where the plot of $S_{\rm ent}$ is almost vertical from the point where it vanishes to where it diverges to $-\infty$.  So to a good approximation, the positivity of geodesic length in the deformed geometry is synonymous with the reality of HEE in the deformed theory.

\begin{figure}[!h]
    \centering
    \includegraphics[width=.68\textwidth]{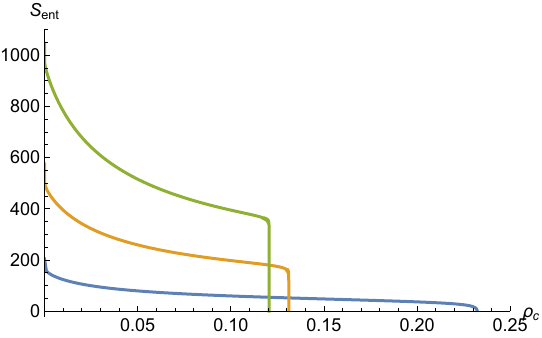}
    \caption{{\small Plot for HEE as obtained from the finite cut-off prescription as a function of the cut-off radius $\rho_c$, for various $l$. Here we have set $\beta=1,\Omega=0.4$. The blue curve corresponds to $l=5,$ the orange curve corresponds to $l=10$, and the green curve corresponds to $l=20$. Note that, as $l$ increases, the maximum allowed value of $\rho_c$ tends to saturate to $\rho_{c,{\rm max}}= - 2 \mu_l$, corresponding to this choice of $\beta$ and $\Omega$.}}
    \label{Fig:FC}
\end{figure}

Since the lower bound puts a cut-off on the negative values of $\mu$, one should expect this feature to show up in the finite cut-off picture as well. Figure \ref{Fig:FC} shows the plot for HEE computed in the finite cut-off picture as a function of the cut-off radius $\rho_c$ for various sizes of the entangling region.

The plots clearly show that, given a fixed temperature and a chemical potential, the cut-off surface can not be put arbitrarily deep inside the bulk while also retaining geodesics between boundary points on a fixed-time slice spacelike. Beyond this upper bound on the cut-off radius, one cannot use the standard definitions of HEE. Also note that when $l\sim \beta$, the maximum allowed value for the cut-off radius seems to be $l$ dependent. However, as we take larger $l$, the upper bound on $\rho_c$ tends to become insensitive to $l$ and is in agreement with the lower bound on (negative) $\mu$, as expected. Note that, here also $S_{\text{ent}}$ is negative in a negligible range of $\rho_c $ before diverging to $ -\infty$ at $\rho_{c,{\rm max}}$. For all the plots in Figure \ref{Fig:FC}, $\rho_{c,{\rm max}} < \rho_h= (L \bar L)^{-1/2}$, so the cut-off remains always outside the horizon\footnote{To check this, one needs to invert \eqref{betaomega} with $\mu=-\rho_c/2$, to express $L$ and $\bar{L}$ in terms of $\beta$, $\Omega$, and $\rho_c $.}.
 
Finally, one might wonder if the lower bound is also reflected in the thermodynamic quantities, like the upper bound. For example, one can compute the thermal entropy (density) from \eqref{entrotbtz} in the high-temperature limit $l\gg\beta_\pm$. This gives,
\begin{align}   \label{eq-thermal-entropy}
S_{\text{thermal}}= \frac{2\pi^2 (\beta_+ +\beta_-)}{\beta_+\beta_-\sqrt{1-\frac{8\pi^2 \mu}{\beta_+\beta_-}}}. 
\end{align}
Clearly, $S_\text{thermal}$ remains finite for any $\mu<\mu_u$. This conclusion remains true if one computes the internal energy or free energy densities as well. It would be rather interesting to look for the imprints of this lower bound on the negative values of the deformation parameter in other contexts of the deformed theory, which we leave for future investigations.



\section{QNEC in deformed theory}\label{Sec4}

In this section, we will study the Quantum Null Energy Condition (QNEC) in the deformed theory. In two dimensions, the QNEC has the following versions depending on whether the theory is conformally invariant or not: 
\begin{subequations}
\begin{align} 
  \text{QFT} &: ~~~~   \mathcal Q_\pm \equiv 2\pi \langle T_{\pm\pm} \rangle - S_{\text{ent}}'' \geq 0 , \label{qftqnec}\\
  \text{CFT}&: ~~~~  \mathcal Q_\pm \equiv 2\pi \langle T_{\pm\pm} \rangle - S_{\text{ent}}'' - \frac{6}{c} S_{\text{ent}}'^2\geq 0. \label{cftqnec}
\end{align}
\end{subequations}
where $T_{\pm\pm}$ are the null components of the stress-tensor and $'$ denotes derivative w.r.t $\partial_\pm$ for $\mathcal{Q}_\pm$, obtained from infinitesimal variations of the entanglement entropy of the interval ending at the point of observation, under shifts along the respective null directions. We are interested in a situation where the undeformed theory is a 1+1 dimensional holographic CFT with an irrelevant deformation breaking the conformal invariance. So in principle, we should consider a QNEC inequality that smoothly interpolates between the two versions \eqref{cftqnec} and \eqref{qftqnec}. However, to the best of our knowledge, such an interpolating version of QNEC does not exist in the literature. At least for small deformations, a naive guess is to propose the following slightly modified version of QNEC:
\begin{equation}    \label{modqnec}
 \mathcal {\tilde{Q}}_\pm \equiv 2\pi \langle T_{\pm\pm} \rangle - S_{\text{ent}}''- \frac{6}{c}S_{\text{ent}}^{(0)'2} \geq 0,
\end{equation}
where $S_{\text{ent}}^{(0)}$ is the entanglement entropy in the undeformed theory.  The last term ensures that at leading order in the deformation, $\mathcal{\tilde Q}_{\pm}$  exactly correspond to the CFT version \eqref{cftqnec}; in particular, it is saturated when the state under consideration is dual to a Banados spacetime \cite{Ecker:2019ocp}. Then for the first correction due to the $T \bar T$ deformation, only the first two terms would contribute. However, we do not have any supporting arguments for this modified version of QNEC. Furthermore, we are interested in the non-perturbative regime of the deformation. So we will simply proceed with the QFT version of the QNEC \eqref{qftqnec} and investigate the inequality as we change the deformation parameter. 

As in the case of entanglement entropy, let us begin with the case of vanishing chemical potential. To evaluate $\mathcal Q_{+(-)}$,  we deform the end-point $p_1= (r_c,0,0)$ (in coordinates $(r,T,X)$ with metric given by \eqref{defbtz}) along $\partial_{U(V)}$ infinitesimally by $\delta$, keeping the other endpoint fixed.  Then we compute the geodesic distance between the points $p_1' =(r_c,+(-)\delta/2, \delta/2)$  and $p_2=(r_c,0,l)$ using the method discussed previously, eventually identifying it with the HEE, which is now a function of $\delta$. Finally,  we take derivatives of this HEE w.r.t $\delta$ followed by setting $\delta=0$ to get $S_{\text{ent}}''$. The stress-tensor components can be read off from \eqref{stress-tensor} as $T_{++}= T_{UU}, T_{--}= T_{VV}$, with $\bar L= L$ and $L$ being given by  \eqref{defL}. Putting all these together,  $\mathcal{Q}_{\pm}$ is evaluated to be 
\begin{align}\label{qnecbtz}
    \mathcal{Q}_+= \mathcal{Q}_-= \frac{c\pi^2 \left(\beta \sqrt{\beta^2 -8\pi^2 \mu}+(\beta^2-4\pi^2 \mu)\csch^2\left(\frac{\pi l}{\sqrt{\beta^2-8\pi^2 \mu}}\right)\right)}{6\beta^2\left(\beta^2-8\pi^2 \mu\right)}.
\end{align}
The expression for $\mathcal Q_{\pm}$ evaluated at the other endpoint $p_2$ is exactly the same as \eqref{qnecbtz}.

\begin{figure}[tbp]
\centering
\begin{subfigure}{.5\textwidth}
  \centering
  \includegraphics[scale=.78]{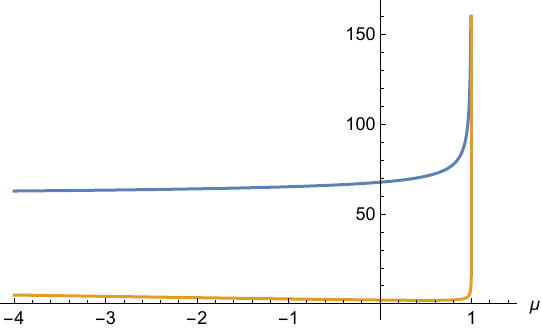}
  \caption{$l=2,\beta = \sqrt{8}\pi,\Omega=0$.}
  \label{Fig:4a}
\end{subfigure}%
\begin{subfigure}{.5\textwidth}
  \centering
  \includegraphics[scale=.78]{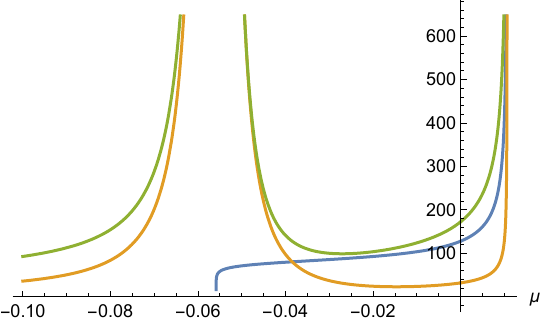}
  \caption{$l=2,\beta=1,\Omega=0.4$.}
  \label{Fig:4b}
\end{subfigure}
\caption{{\small \textbf{Left}: $\mathcal{Q}_{\pm}$ (orange curve)   plotted along with $S_{\text{ent}}$ (blue curve) as a function of the deformation parameter $\mu$ for vanishing $\Omega$, with the other parameters specified in the subcaption. The upper bound is at $\mu_u=\beta^2/8\pi^2=1$, beyond which  $S_{\text{ent}}$ and $\mathcal{Q}_{\pm}$ all become complex. Below this bound, they remain to be real and non-negative. \textbf{Right}: Plot for $\mathcal{Q}_+$ (orange curve), $\mathcal{Q}_-$ (green curve) along with $S_{\text{ent}}$ (blue curve)  as a function of the deformation parameter in the presence of a finite chemical potential, with the values of various parameters specified in the subcaption. Note that, in this case as well, for $\mu >\mu_u$, all three quantities become complex. But now there is also a lower bound at $\mu_l \simeq -0.055$, where these quantities become complex again. However, for $\mu < \mu_l$, $S_{\text{ent}}$ remains to be complex whereas $\mathcal{Q}_{\pm}$ become real and non-negative.}}
\label{Fig:4}
\end{figure}

A small $\mu$ expansion of \eqref{qnecbtz} gives
\begin{align} \label{qnecbtzser}
\mathcal{Q}_{\pm}= \frac{c\pi^2}{6\beta^2}\coth^2\left(\frac{\pi l}{\beta}\right)+ \frac{c\pi^4}{3\beta^4}\coth\left(\frac{\pi l}{\beta}\right)\csch^2\left(\frac{\pi l}{\beta}\right)\left(-4\left(\frac{\pi l}{\beta}\right)+\sinh\left(\frac{2\pi l}{\beta}\right)\right)\mu +\mathcal{O}(\mu^2).
\end{align}
The leading term is simply equal to $\frac{6}{c}S_{\text{ent}}^{(0)'2}$, which is expected as we are working with the QFT version of QNEC \eqref{qftqnec}. Said differently, $\mathcal{Q}_{\pm}=0$ in the state dual to the undeformed non-rotating planar BTZ, for which the CFT version \eqref{cftqnec} must be considered. From \eqref{qnecbtz}, it is also clear that $\mathcal{Q}_{\pm}$ become complex beyond the same upper bound \eqref{ubound}. However, for negative $\mu$, $\mathcal{Q}_{\pm} >0$. Figure \ref{Fig:4a} shows the variation of $\mathcal{Q}_{\pm}$ (along with $S_{\text{ent}}$) with the deformation for fixed values of $\beta$ and $l$. Clearly, QNEC is never violated in the deformed theory as long as $\mu < \mu_u$.

Next, we consider the case of a finite chemical potential  $\Omega$. We shall proceed similarly by shifting the endpoint  $p_1=(r_c,0,0)$ in the $(r,T,X)$ coordinates (in which the metric is given by\eqref{defrotbtz}) to $p_1'=(r_c, \pm \delta/2, \delta/2)$ while keeping the other endpoint fixed at $p_2=(r_c,0,l)$. In this case, $\bar L \neq L$, leading to $\mathcal{Q}_+ \neq \mathcal{Q}_-$. The exact expressions for $\mathcal Q_\pm$ are not very illuminating so we refrain from reporting them. In the small $\mu$ limit, we have
\begin{align}
    \mathcal{Q_+}=&\ \frac{c\pi^2}{6\beta_+^2}\coth^2\left(\frac{\pi l}{\beta_+}\right) 
    +\frac{c\pi^4}{3\beta_+^3 \beta_-}\coth\left(\frac{\pi l}{\beta_+}\right)\csch^2\left(\frac{\pi l}{\beta_+}\right)\left(-2\pi l\left(\frac{1}{\beta_+}+\frac{1}{\beta_-}\right)+\sinh\left(\frac{2\pi l}{\beta_+}\right)\right)\mu +\mathcal{O}(\mu^2), \nonumber \\ 
    \mathcal{Q_-}=&\ \frac{c\pi^2}{6\beta_-^2}\coth^2\left(\frac{\pi l}{\beta_-}\right) 
    + \frac{c\pi^4}{3\beta_+ \beta_-^3}\coth\left(\frac{\pi l}{\beta_-}\right)\csch^2\left(\frac{\pi l}{\beta_-}\right)\left(-2\pi l\left(\frac{1}{\beta_+}+\frac{1}{\beta_-}\right)+\sinh\left(\frac{2\pi l}{\beta_-}\right)\right)\mu +\mathcal{O}(\mu^2). \nonumber
\end{align}
Clearly, they agree with \eqref{qnecbtzser} in the limit $\beta_-\rightarrow \beta_+ = \beta$. 

Figure \ref{Fig:4b} shows the variation of the non-perturbative $\mathcal{Q}_{\pm}$ (along with $S_{\text{ent}}$) with the deformation parameter for fixed values of $\beta, \Omega$ and $l$. Note that there is a qualitative difference in the behaviour of $S_{\text{ent}}$ and $\mathcal{Q}_{\pm}$ across the two bounds. In case of the upper bound, all the three quantities become complex when $\mu > \mu_{u}$. However, in case of the lower bound, when $\mu < \mu_{l}$, $S_{\text{ent}}$ becomes complex whereas $\mathcal{Q}_{\pm}$ remains real and non-negative in the range $-\infty <\mu <\mu_{l}$. This behaviour is not difficult to understand if we recall how the two bounds were derived. The upper bound corresponds to the value of $\mu$ when the expression appearing under the square roots in \eqref{entrotbtz} vanishes. The derivatives of $S_{\text{ent}}$ involved in $\mathcal{Q}_{\pm}$ do not change the expression under the square root. So $\mathcal{Q_{\pm}},S_{\text{ent}}$ all become complex beyond this upper bound. However, the lower bound was derived from the zero of the argument of the logarithmic function in \eqref{entrotbtz}. Clearly, while taking derivatives, new terms are expected to appear, and there is a priori no reason for  $S_{\text{ent}}^{''}$ to respect the same lower bound which $S_{\text{ent}}$ does. However, quite surprisingly, the lower bound still appears to be a special (isolated) point where $S_{\text{ent}}^{''}$ and hence $\mathcal{Q}_{\pm}$  become complex. Below this bound, it again becomes real and non-negative, but since $S_{\text{ent}}$  remains to be complex, reality or positivity of $\mathcal{Q}_{\pm}$ do not seem to make much sense. However, within the allowed range of the deformation parameter, QNEC continues to be satisfied even in the deformed theory.

\section{Conclusions and Discussion}\label{Sec5}

In this paper, we have studied the holographic entanglement entropy and quantum null energy condition in finite temperature states of a $T\bar{T}$ deformed 1+1-dimensional holographic CFT, both in the presence and absence of a chemical potential for boost. Our holographic computations are based on the mixed boundary condition interpretation of the deformation. First, we obtained the relevant deformed spacetimes from the corresponding undeformed ones using the state-dependent coordinate transformations \eqref{pcfmap2}. These deformed spacetimes are locally AdS$_3$, so they can be mapped to the Poincare patch where a closed-form expression is known for geodesic distances between arbitrary points. Using this algorithm, we computed the HEE and evaluated the QNEC inequality in the deformed theory.

In the case of a vanishing chemical potential, our (non-perturbative) result for the HEE agrees with various known results in the literature, including the upper bound on the deformation parameter which follows from the the reality of the HEE in the deformed theory. This upper bound is analogous to the one that follows from the reality of the ground state energy in a deformed CFT on a cylinder, with the size of the cylinder $R$ replacing the size of the thermal circle $\beta$. In fact, this upper bound is nothing but the Hagedorn bound, beyond which thermal equilibrium fails to make sense in the deformed theory.  

In the presence of a finite chemical potential associated with a boost, our analysis reveals a new lower bound on the negative values of the deformation parameter, thus confining its allowed values within the window \eqref{murange} for the sake of reality of HEE in the deformed theory or equivalently, the positivity of geodesic lengths in the deformed geometry. However, this lower bound does not show up in the thermal entropy or any other thermodynamic quantities, so its physical interpretation is not very clear to us otherwise.  From a holographic perspective, this lower bound turns out to be related to a spacelike to null transition of the associated RT surface in the deformed geometry and this requires the presence of a finite chemical potential along with the deformation. Like the upper bound, this lower bound is also independent of the size of the entangling region, being completely fixed by the temperature and the chemical potential. During the course of our computations, we have also demonstrated that our results agree with the ones obtained in the finite cut-off approach only in the high-temperature limit.

We have also explored the QNEC inequality in these states, based on the minimal assumption of the existence of a stress-tensor and entanglement entropy in the deformed theory. Our computations show that the QNEC inequality remains satisfied in the deformed theory for both positive as well as negative values of the deformation parameter, within the allowed regime. 

Let us comment on the significance of our lower bound. The transition from spacelike to null RT surfaces at the lower bound seems to suggest that the holographic prescription for calculating entanglement entropy in the deformed theory becomes ill-defined at the lower bound, at least for finite entangling regions. A more severe possibility is that entanglement entropy itself is not well-defined at finite length scales in the deformed theory for $\mu \leq \mu_l$. A non-perturbative field-theoretic computation of the entanglement entropy in boosted thermal states of the deformed theory will be essential to clarify this. In contrast, the thermal entropy obtained from the $l \gg \beta_{\pm}$ limit of the holographic entanglement entropy \eqref{entrotbtz} remains real below this lower bound. These two observations together perhaps indicate the non-local nature of the deformed theory. However, this lower bound is qualitatively different from the Hagedorn bound, in the sense that it exists for the negative sign of the deformation parameter, and also the thermodynamic quantities are not sensitive to it.

It will be interesting to further investigate if this lower bound shows up in other aspects of the deformed theory, for a deeper understanding of its significance. For instance, one can calculate Renyi entropies for states dual to the deformed boosted BTZ geometries and check its reality properties at the bounds that follow from the reality of HEE.  Also, our study should be extended to generic states of the deformed theory which are dual to the deformed spacetimes with non-constant $L$ and $\bar L$.  However, in that case, the state-dependent transformations \eqref{pcfmap1} may not be invertible, and hence analytic computations are beyond the scope of this paper. Nevertheless, it would be interesting to numerically explore if there are further constraints on the deformation parameter based on the reality of the HEE in such states.  Also, given that QNEC remains satisfied in the deformed theory, one can further study the Quantum Dominant Energy Condition (QDEC) \cite{Wall:2017blw} in these states, which involves the entire stress-tensor rather than just the null components. However, since QDEC is not well-established at this point even in the undeformed CFT, we postpone this for future work.

\begin{acknowledgments}
We would like to thank Giuseppe Policastro and Ayan Mukhopadhyay for their insightful comments and their collaboration during the initial stages of this work. We also thank Soumangsu Chakraborty, Kausik Ghosh, Sam van Leuven, and Aron Wall for useful discussions.  Research of AB is partially supported by the European  MSCA grant HORIZON-MSCA-2022-PF-01-01 and by the H.F.R.I call “Basic research Financing (Horizontal support of all Sciences)” under the National Recovery and Resilience Plan “Greece 2.0” funded by the European Union – NextGenerationEU (H.F.R.I. Project Number: 15384). The research of PR is supported by the South African Research Chairs Initiative of the Department of Science and Technology and by the National Research Foundation.  PR would also like to thank the Isaac Newton Institute for Mathematical Sciences for support and hospitality during the programme “Black holes: bridges between number theory and holographic quantum information” where part of this work was done. The programme was supported by EPSRC Grant Number EP/R014604/1.

\end{acknowledgments}

\appendix

\section{Some relevant identities} \label{Id}
In this Appendix we explicitly work out certain relations that lead to (\ref{solfe1})-(\ref{solfe2}) starting from (\ref{fe2}). With the redefinition $\hat{T}_{\al\bt}= T_{\al\bt} - \gm_{\al\bt} T$, (\ref{fe21}) follows trivially whereas to get (\ref{fe22}) we first note that
\begin{align} \label{Id1}
    \partial_{\mu}\sqrt{\gm}=\frac{1}{2}\sqrt{\gm} \gm^{\al\bt}\partial_{\mu}\gm_{\al\bt}=- \sqrt{\gm} \gm^{\al\bt}\left(T_{\al\bt}-\gm_{\al\bt} T\right) = \sqrt{\gm} T .
\end{align}
With this identity the second equation of (\ref{fe2}) now leads to
\begin{equation}\label{Id2}
   \partial_{\mu}T_{\al\bt}=-\gm_{\al\bt} \mathcal{O}_{T\bar{T}} - T_{\al\gm}T_{\bt}^{\gm}. 
\end{equation}
This gives
\begin{align}\label{Id3}
   \partial_{\mu}\hat{T}_{\al\bt}=& \partial_{\mu}\left(T_{\al\bt} - \gm_{\al\bt} T\right)= -\gm_{\al\bt} \mathcal{O}_{T\bar{T}} - T_{\al\gm}T_{\bt}^{\gm}+ 2T \left(T_{\al\bt} - \gm_{\al\bt} T\right) -\gm_{\al\bt}\partial_{\mu}T.
   \end{align}
Now,
\begin{align}
\partial_{\mu} T =& \left(\partial_{\mu}\gm^{\al\bt}\right)T_{\al\bt} + \gm^{\al\bt}\partial_{\mu}T_{\al\bt} \nonumber\\
=& 2 \left(T^{\al\bt}-\gm^{\al\bt} T\right)T_{\al\bt}+ \gm^{\al\bt}\left(-\gm_{\al\bt} \mathcal{O}_{T\bar{T}} - T_{\al\gm}T_{\bt}^{\gm}\right) \nonumber\\
=& 2 T^{\al\bt}T_{\al\bt} - 2T^2 - 2 \mathcal{O}_{T\bar{T}}-T^{\al\bt}T_{\al\bt} \nonumber\\
=& -T^{\delta\sigma}T_{\delta\sigma}.
\end{align}
Putting this back in (\ref{Id3}) and using the definition of $\mathcal{O}_{T\bar{T}}$ we have 
\begin{align}\label{Id4}
   \partial_{\mu}\hat{T}_{\al\bt}=  - T_{\al\gm}T_{\bt}^{\gm}+ 2T T_{\al\bt} - \gm_{\al\bt} T^2=   -\hat{T}_{\al\gm}\hat{T}_{\bt}^{\gm},
   \end{align}
   which is (\ref{fe22}). Finally note that,
 \begin{align}
\partial_{\mu} \left(\hat{T}_{\al\gm}\hat{T}_{\bt}^{\gm}\right) =& \left(\partial_{\mu}\hat{T}_{\al\gm}\right)\hat{T}_{\bt}^{\gm} + \hat{T}_{\al\gm}\left(\partial_{\mu}\hat{T}_{\bt\delta}\right) \gm^{\delta\gm}+ \hat{T}_{\al\gm}\hat{T}_{\bt\delta} \partial_{\mu}{\gm}^{\delta\gm} \nonumber\\
=& -2 \hat{T}_{\al\gm}\hat{T}_{\bt\delta} \hat{T}^{\gm\delta}+ 2 \hat{T}_{\al\gm}\hat{T}_{\bt\delta}\left(T^{\gm\delta}-\gm^{\gm\delta}T\right)=0. 
\end{align}  
So $\hat{T}_{\al\gm}T_{\bt}^{\gm}$ is invariant along the flow. Eqns (\ref{fe31}-\ref{fe32}) supplemented with this identity finally lead to (\ref{solfe1}-\ref{solfe2}).
\section{Coordinate transformations} \label{maps}
In this Appendix, we provide the details of the coordinate transformations that take the deformed Banados spacetimes (\ref{defbtz}) and (\ref{defrotbtz}) to the Poincare patch (\ref{Poincare}).

First, consider the case of deformed non-rotating  BTZ. Starting from (\ref{defbtz}) we first absorb the deformation dependent factors to define the scaled  time and spatial coordinates 
\begin{equation} \label{defbtzrescale}
\tilde{\mathcal T}= \frac{T}{1+2\mu L} ,\qquad \mathcal{X}= \frac{ X}{1-2 \mu L} \ .
\end{equation}
Note that in ($r,\tilde{\mathcal{T}},\mathcal{X}$) coordinates, (\ref{defbtz}) exactly takes the form of an undeformed BTZ,
\begin{equation}\label{A11}
    ds^2= \frac{dr^2}{r^2-r_h^2} - (r^2-r_h^2) d\tilde{\mathcal{T}}^2  + r^2~d\mathcal{X}^2 .
\end{equation}
Next we define
\[
d\mathcal T=d\tilde{\mathcal T}+dr_*\ ,\qquad dr_*= \frac{dr}{r^2-r_h^2},\qquad r=1/z,
\]
to put  (\ref{A11}) to the ingoing Eddington-Finkelstein (EF) form,
\begin{equation}		\label{defbtzef}
ds^2= -\frac{2}{z^2} d\mathcal T\, dz -\left(\frac{1}{z^2} - r_h^2\right)d\mathcal T^2 + \frac{1}{z^2} d\mathcal{X}^2.
\end{equation}
The finite coordinate transformations are
\begin{equation}
    {\mathcal T} = \tilde{\mathcal T} + r_*, \qquad r_* = - \frac{1}{r_h}\arctanh{\left(\frac{r_h}{r}\right)} ,
\end{equation}
and the range of the radial coordinate $z$ is now $ z_c \leq z\leq z_h$, with $z_c = 1/{r_c} \rightarrow 0$ and $z_h=1/{r_h}.$
 Finally we put this metric to the Poincare EF form 
\begin{equation}            \label{eq:metric_Poincare}
ds^2=\frac{-2 dZ dW -dW^2 +dY^2}{Z^2}
\end{equation}
with the transformation
\begin{align} 
Z= e^{r_h \mathcal T} \frac{r_h z}{1-r_h z},\qquad
W= e^{r_h \mathcal T} \frac{\cosh(r_h \mathcal{X}) -r_h z}{1-r_h z},\qquad 
Y= e^{r_h \mathcal T} \frac{\sinh r_h \mathcal{X} }{1-r_h z}.	\label{EtoP}
\end{align}
The inverse map is given by 
\begin{align}
z=&\ \frac{1}{r_h}\frac{Z}{\sqrt{(Z+W)^2-Y^2}},\nonumber\\
\mathcal T=&\ \frac{1}{r_h}\log\left[\sqrt{(Z+W)^2-Y^2}-Z\right],\\
\mathcal{X}=&\ \frac{1}{2r_h}\log\left[\frac{Z+W+Y}{Z+W-Y}\right]. \nonumber
\end{align}
In the case of the deformed boosted BTZ, the set of coordinate transformations is analogous. Starting from (\ref{defrotbtz}), we first define the coordinates $(R,\mathcal{\tilde T},\mathcal{X})$ via
\begin{align}
    R = \sqrt{r^2-r_-^2},\quad \mathcal{\tilde T} = \frac{2(r_+T-r_-X)}{r_h(2+r_h^2\mu)}, \quad \mathcal{X} = \frac{2(r_+X-r_-T)}{r_h(2-r_h^2\mu)},
\end{align}
so that (\ref{defrotbtz}) takes the form of (\ref{A11}) with $r$ replaced by $R$ and $r_h^2 = r_+^2-r_-^2$.  The remaining steps to put this metric to the Poincare patch are exactly the same as in the non-rotating case.

In order to compute $S_{\text{ent}}$ and $\mathcal{Q}_{\pm}$, we first choose the endpoints of the geodesic in the $(r,T,X)$ coordinates of \eqref{defbtz} or \eqref{defrotbtz} and use the transformations mentioned above to map the endpoints to the Poincare patch $(Z,W,Y)$. Then we use \eqref{invariant}  and \eqref{length} to compute the geodesic length.



\bibliography{refs}{}

\providecommand{\href}[2]{#2}\begingroup\raggedright\begin{thebibliography}{10}

\bibitem{Smirnov:2016lqw}
F.A.~Smirnov and A.B.~Zamolodchikov, \emph{{On space of integrable quantum
  field theories}},
  \href{https://doi.org/10.1016/j.nuclphysb.2016.12.014}{\emph{Nucl. Phys. B}
  {\bfseries 915} (2017) 363}
  [\href{https://arxiv.org/abs/1608.05499}{{\ttfamily 1608.05499}}].

\bibitem{Dubovsky:2012wk}
S.~Dubovsky, R.~Flauger and V.~Gorbenko, \emph{{Solving the Simplest Theory of
  Quantum Gravity}}, \href{https://doi.org/10.1007/JHEP09(2012)133}{\emph{JHEP}
  {\bfseries 09} (2012) 133} [\href{https://arxiv.org/abs/1205.6805}{{\ttfamily
  1205.6805}}].

\bibitem{Zamolodchikov:2004ce}
A.B.~Zamolodchikov, \emph{{Expectation value of composite field T anti-T in
  two-dimensional quantum field theory}},
  \href{https://arxiv.org/abs/hep-th/0401146}{{\ttfamily hep-th/0401146}}.

\bibitem{Cardy:2018sdv}
J.~Cardy, \emph{{The $ T\overline{T} $ deformation of quantum field theory as
  random geometry}}, \href{https://doi.org/10.1007/JHEP10(2018)186}{\emph{JHEP}
  {\bfseries 10} (2018) 186}
  [\href{https://arxiv.org/abs/1801.06895}{{\ttfamily 1801.06895}}].

\bibitem{Datta:2018thy}
S.~Datta and Y.~Jiang, \emph{{$T\bar{T}$ deformed partition functions}},
  \href{https://doi.org/10.1007/JHEP08(2018)106}{\emph{JHEP} {\bfseries 08}
  (2018) 106} [\href{https://arxiv.org/abs/1806.07426}{{\ttfamily
  1806.07426}}].

\bibitem{Aharony:2018bad}
O.~Aharony, S.~Datta, A.~Giveon, Y.~Jiang and D.~Kutasov, \emph{{Modular
  invariance and uniqueness of $T\bar{T}$ deformed CFT}},
  \href{https://doi.org/10.1007/JHEP01(2019)086}{\emph{JHEP} {\bfseries 01}
  (2019) 086} [\href{https://arxiv.org/abs/1808.02492}{{\ttfamily
  1808.02492}}].

\bibitem{Callebaut:2019omt}
N.~Callebaut, J.~Kruthoff and H.~Verlinde, \emph{{$ T\overline{T} $ deformed
  CFT as a non-critical string}},
  \href{https://doi.org/10.1007/JHEP04(2020)084}{\emph{JHEP} {\bfseries 04}
  (2020) 084} [\href{https://arxiv.org/abs/1910.13578}{{\ttfamily
  1910.13578}}].

\bibitem{Tolley:2019nmm}
A.J.~Tolley, \emph{{$ T\overline{T} $ deformations, massive gravity and
  non-critical strings}},
  \href{https://doi.org/10.1007/JHEP06(2020)050}{\emph{JHEP} {\bfseries 06}
  (2020) 050} [\href{https://arxiv.org/abs/1911.06142}{{\ttfamily
  1911.06142}}].

\bibitem{Dubovsky:2017cnj}
S.~Dubovsky, V.~Gorbenko and M.~Mirbabayi, \emph{{Asymptotic fragility, near
  AdS$_{2}$ holography and $ T\overline{T} $}},
  \href{https://doi.org/10.1007/JHEP09(2017)136}{\emph{JHEP} {\bfseries 09}
  (2017) 136} [\href{https://arxiv.org/abs/1706.06604}{{\ttfamily
  1706.06604}}].

\bibitem{Dubovsky:2018bmo}
S.~Dubovsky, V.~Gorbenko and G.~Hern\'andez-Chifflet, \emph{{$ T\overline{T} $
  partition function from topological gravity}},
  \href{https://doi.org/10.1007/JHEP09(2018)158}{\emph{JHEP} {\bfseries 09}
  (2018) 158} [\href{https://arxiv.org/abs/1805.07386}{{\ttfamily
  1805.07386}}].

\bibitem{McGough:2016lol}
L.~McGough, M.~Mezei and H.~Verlinde, \emph{{Moving the CFT into the bulk with
  $ T\overline{T} $}},
  \href{https://doi.org/10.1007/JHEP04(2018)010}{\emph{JHEP} {\bfseries 04}
  (2018) 010} [\href{https://arxiv.org/abs/1611.03470}{{\ttfamily
  1611.03470}}].

\bibitem{Kraus:2018xrn}
P.~Kraus, J.~Liu and D.~Marolf, \emph{{Cutoff AdS$_{3}$ versus the $
  T\overline{T} $ deformation}},
  \href{https://doi.org/10.1007/JHEP07(2018)027}{\emph{JHEP} {\bfseries 07}
  (2018) 027} [\href{https://arxiv.org/abs/1801.02714}{{\ttfamily
  1801.02714}}].

\bibitem{Taylor:2018xcy}
M.~Taylor, \emph{{TT deformations in general dimensions}},
  \href{https://arxiv.org/abs/1805.10287}{{\ttfamily 1805.10287}}.

\bibitem{Kraus:2022mnu}
P.~Kraus, R.~Monten and K.~Roumpedakis, \emph{{Refining the cutoff 3d gravity/$
  T\overline{T} $ correspondence}},
  \href{https://doi.org/10.1007/JHEP10(2022)094}{\emph{JHEP} {\bfseries 10}
  (2022) 094} [\href{https://arxiv.org/abs/2206.00674}{{\ttfamily
  2206.00674}}].

\bibitem{Guica:2019nzm}
M.~Guica and R.~Monten, \emph{{$T\bar T$ and the mirage of a bulk cutoff}},
  \href{https://doi.org/10.21468/SciPostPhys.10.2.024}{\emph{SciPost Phys.}
  {\bfseries 10} (2021) 024}
  [\href{https://arxiv.org/abs/1906.11251}{{\ttfamily 1906.11251}}].

\bibitem{Jiang:2019epa}
Y.~Jiang, \emph{{A pedagogical review on solvable irrelevant deformations of 2D
  quantum field theory}},
  \href{https://doi.org/10.1088/1572-9494/abe4c9}{\emph{Commun. Theor. Phys.}
  {\bfseries 73} (2021) 057201}
  [\href{https://arxiv.org/abs/1904.13376}{{\ttfamily 1904.13376}}].

\bibitem{Guica:2022abc}
M.~Guica, ``{$T \bar T$ deformations and holography}.''
  \url{https://indico.cern.ch/event/857396/contributions/3706292/attachments/2036750/3410352/ttbar_cern_v1s.pdf}.

\bibitem{Donnelly:2018bef}
W.~Donnelly and V.~Shyam, \emph{{Entanglement entropy and $T \overline{T}$
  deformation}},
  \href{https://doi.org/10.1103/PhysRevLett.121.131602}{\emph{Phys. Rev. Lett.}
  {\bfseries 121} (2018) 131602}
  [\href{https://arxiv.org/abs/1806.07444}{{\ttfamily 1806.07444}}].

\bibitem{Chen:2018eqk}
B.~Chen, L.~Chen and P.-X.~Hao, \emph{{Entanglement entropy in
  $T\overline{T}$-deformed CFT}},
  \href{https://doi.org/10.1103/PhysRevD.98.086025}{\emph{Phys. Rev. D}
  {\bfseries 98} (2018) 086025}
  [\href{https://arxiv.org/abs/1807.08293}{{\ttfamily 1807.08293}}].

\bibitem{Gorbenko:2018oov}
V.~Gorbenko, E.~Silverstein and G.~Torroba, \emph{{dS/dS and $ T\overline{T}
  $}}, \href{https://doi.org/10.1007/JHEP03(2019)085}{\emph{JHEP} {\bfseries
  03} (2019) 085} [\href{https://arxiv.org/abs/1811.07965}{{\ttfamily
  1811.07965}}].

\bibitem{Murdia:2019fax}
C.~Murdia, Y.~Nomura, P.~Rath and N.~Salzetta, \emph{{Comments on holographic
  entanglement entropy in $TT$ deformed conformal field theories}},
  \href{https://doi.org/10.1103/PhysRevD.100.026011}{\emph{Phys. Rev. D}
  {\bfseries 100} (2019) 026011}
  [\href{https://arxiv.org/abs/1904.04408}{{\ttfamily 1904.04408}}].

\bibitem{Ota:2019yfe}
T.~Ota, \emph{{Comments on holographic entanglements in cutoff AdS}},
  \href{https://arxiv.org/abs/1904.06930}{{\ttfamily 1904.06930}}.

\bibitem{Banerjee:2019ewu}
A.~Banerjee, A.~Bhattacharyya and S.~Chakraborty, \emph{{Entanglement Entropy
  for $TT$ deformed CFT in general dimensions}},
  \href{https://doi.org/10.1016/j.nuclphysb.2019.114775}{\emph{Nucl. Phys. B}
  {\bfseries 948} (2019) 114775}
  [\href{https://arxiv.org/abs/1904.00716}{{\ttfamily 1904.00716}}].

\bibitem{Jeong:2019ylz}
H.-S.~Jeong, K.-Y.~Kim and M.~Nishida, \emph{{Entanglement and R\'enyi entropy
  of multiple intervals in $T\overline{T}$-deformed CFT and holography}},
  \href{https://doi.org/10.1103/PhysRevD.100.106015}{\emph{Phys. Rev. D}
  {\bfseries 100} (2019) 106015}
  [\href{https://arxiv.org/abs/1906.03894}{{\ttfamily 1906.03894}}].

\bibitem{He:2019vzf}
S.~He and H.~Shu, \emph{{Correlation functions, entanglement and chaos in the $
  T\overline{T}/J\overline{T} $-deformed CFTs}},
  \href{https://doi.org/10.1007/JHEP02(2020)088}{\emph{JHEP} {\bfseries 02}
  (2020) 088} [\href{https://arxiv.org/abs/1907.12603}{{\ttfamily
  1907.12603}}].

\bibitem{Donnelly:2019pie}
W.~Donnelly, E.~LePage, Y.-Y.~Li, A.~Pereira and V.~Shyam, \emph{{Quantum
  corrections to finite radius holography and holographic entanglement
  entropy}}, \href{https://doi.org/10.1007/JHEP05(2020)006}{\emph{JHEP}
  {\bfseries 05} (2020) 006}
  [\href{https://arxiv.org/abs/1909.11402}{{\ttfamily 1909.11402}}].

\bibitem{Asrat:2020uib}
M.~Asrat and J.~Kudler-Flam, \emph{{$T\bar{T}$, the entanglement wedge cross
  section, and the breakdown of the split property}},
  \href{https://doi.org/10.1103/PhysRevD.102.045009}{\emph{Phys. Rev. D}
  {\bfseries 102} (2020) 045009}
  [\href{https://arxiv.org/abs/2005.08972}{{\ttfamily 2005.08972}}].

\bibitem{Allameh:2021moy}
K.~Allameh, A.F.~Astaneh and A.~Hassanzadeh, \emph{{Aspects of holographic
  entanglement entropy for TT\textasciimacron{}-deformed CFTs}},
  \href{https://doi.org/10.1016/j.physletb.2022.136914}{\emph{Phys. Lett. B}
  {\bfseries 826} (2022) 136914}
  [\href{https://arxiv.org/abs/2111.11338}{{\ttfamily 2111.11338}}].

\bibitem{Setare:2022qls}
M.R.~Setare and S.N.~Sajadi, \emph{{Holographic entanglement entropy in
  $T{\bar{T}}$-deformed CFTs}},
  \href{https://doi.org/10.1007/s10714-022-02971-y}{\emph{Gen. Rel. Grav.}
  {\bfseries 54} (2022) 85} [\href{https://arxiv.org/abs/2203.16445}{{\ttfamily
  2203.16445}}].

\bibitem{He:2022xkh}
S.~He, Z.-C.~Liu and Y.~Sun, \emph{{Entanglement entropy and modular
  Hamiltonian of free fermion with deformations on a torus}},
  \href{https://doi.org/10.1007/JHEP09(2022)247}{\emph{JHEP} {\bfseries 09}
  (2022) 247} [\href{https://arxiv.org/abs/2207.06308}{{\ttfamily
  2207.06308}}].

\bibitem{Jeong:2022jmp}
H.-S.~Jeong, W.-B.~Pan, Y.-W.~Sun and Y.-T.~Wang, \emph{{Holographic study of $
  T\overline{T} $ like deformed HV QFTs: holographic entanglement entropy}},
  \href{https://doi.org/10.1007/JHEP02(2023)018}{\emph{JHEP} {\bfseries 02}
  (2023) 018} [\href{https://arxiv.org/abs/2211.00518}{{\ttfamily
  2211.00518}}].

\bibitem{He:2023xnb}
M.~He and Y.~Sun, \emph{{Holographic entanglement entropy in TT-deformed
  AdS3}}, \href{https://doi.org/10.1016/j.nuclphysb.2023.116190}{\emph{Nucl.
  Phys. B} {\bfseries 990} (2023) 116190}
  [\href{https://arxiv.org/abs/2301.04435}{{\ttfamily 2301.04435}}].

\bibitem{Ryu:2006bv}
S.~Ryu and T.~Takayanagi, \emph{{Holographic derivation of entanglement entropy
  from AdS/CFT}},
  \href{https://doi.org/10.1103/PhysRevLett.96.181602}{\emph{Phys. Rev. Lett.}
  {\bfseries 96} (2006) 181602}
  [\href{https://arxiv.org/abs/hep-th/0603001}{{\ttfamily hep-th/0603001}}].

\bibitem{Hubeny:2007xt}
V.E.~Hubeny, M.~Rangamani and T.~Takayanagi, \emph{{A Covariant holographic
  entanglement entropy proposal}},
  \href{https://doi.org/10.1088/1126-6708/2007/07/062}{\emph{JHEP} {\bfseries
  07} (2007) 062} [\href{https://arxiv.org/abs/0705.0016}{{\ttfamily
  0705.0016}}].

\bibitem{Bousso:2015mna}
R.~Bousso, Z.~Fisher, S.~Leichenauer and A.C.~Wall, \emph{{Quantum focusing
  conjecture}}, \href{https://doi.org/10.1103/PhysRevD.93.064044}{\emph{Phys.
  Rev. D} {\bfseries 93} (2016) 064044}
  [\href{https://arxiv.org/abs/1506.02669}{{\ttfamily 1506.02669}}].

\bibitem{Bousso:2015wca}
R.~Bousso, Z.~Fisher, J.~Koeller, S.~Leichenauer and A.C.~Wall, \emph{{Proof of
  the Quantum Null Energy Condition}},
  \href{https://doi.org/10.1103/PhysRevD.93.024017}{\emph{Phys. Rev. D}
  {\bfseries 93} (2016) 024017}
  [\href{https://arxiv.org/abs/1509.02542}{{\ttfamily 1509.02542}}].

\bibitem{Malik:2019dpg}
T.A.~Malik and R.~Lopez-Mobilia, \emph{{Proof of the quantum null energy
  condition for free fermionic field theories}},
  \href{https://doi.org/10.1103/PhysRevD.101.066028}{\emph{Phys. Rev. D}
  {\bfseries 101} (2020) 066028}
  [\href{https://arxiv.org/abs/1910.07594}{{\ttfamily 1910.07594}}].

\bibitem{Koeller:2015qmn}
J.~Koeller and S.~Leichenauer, \emph{{Holographic Proof of the Quantum Null
  Energy Condition}},
  \href{https://doi.org/10.1103/PhysRevD.94.024026}{\emph{Phys. Rev. D}
  {\bfseries 94} (2016) 024026}
  [\href{https://arxiv.org/abs/1512.06109}{{\ttfamily 1512.06109}}].

\bibitem{Balakrishnan:2017bjg}
S.~Balakrishnan, T.~Faulkner, Z.U.~Khandker and H.~Wang, \emph{{A General Proof
  of the Quantum Null Energy Condition}},
  \href{https://doi.org/10.1007/JHEP09(2019)020}{\emph{JHEP} {\bfseries 09}
  (2019) 020} [\href{https://arxiv.org/abs/1706.09432}{{\ttfamily
  1706.09432}}].

\bibitem{Ceyhan:2018zfg}
F.~Ceyhan and T.~Faulkner, \emph{{Recovering the QNEC from the ANEC}},
  \href{https://doi.org/10.1007/s00220-020-03751-y}{\emph{Commun. Math. Phys.}
  {\bfseries 377} (2020) 999}
  [\href{https://arxiv.org/abs/1812.04683}{{\ttfamily 1812.04683}}].

\bibitem{Kudler-Flam:2023hkl}
J.~Kudler-Flam, S.~Leutheusser, A.A.~Rahman, G.~Satishchandran and
  A.J.~Speranza, \emph{{A covariant regulator for entanglement entropy: proofs
  of the Bekenstein bound and QNEC}},
  \href{https://arxiv.org/abs/2312.07646}{{\ttfamily 2312.07646}}.

\bibitem{Lashkari:2018nsl}
N.~Lashkari, \emph{{Constraining Quantum Fields using Modular Theory}},
  \href{https://doi.org/10.1007/JHEP01(2019)059}{\emph{JHEP} {\bfseries 01}
  (2019) 059} [\href{https://arxiv.org/abs/1810.09306}{{\ttfamily
  1810.09306}}].

\bibitem{Moosa:2020jwt}
M.~Moosa, P.~Rath and V.P.~Su, \emph{{A R\'enyi quantum null energy condition:
  proof for free field theories}},
  \href{https://doi.org/10.1007/JHEP01(2021)064}{\emph{JHEP} {\bfseries 01}
  (2021) 064} [\href{https://arxiv.org/abs/2007.15025}{{\ttfamily
  2007.15025}}].

\bibitem{Roy:2022yzm}
P.~Roy, \emph{{Proof of the R\'enyi quantum null energy condition for free
  fermions}}, \href{https://doi.org/10.1103/PhysRevD.108.045010}{\emph{Phys.
  Rev. D} {\bfseries 108} (2023) 045010}
  [\href{https://arxiv.org/abs/2212.02331}{{\ttfamily 2212.02331}}].

\bibitem{Kibe:2021qjy}
T.~Kibe, A.~Mukhopadhyay and P.~Roy, \emph{{Quantum Thermodynamics of
  Holographic Quenches and Bounds on the Growth of Entanglement from the
  Quantum Null Energy Condition}},
  \href{https://doi.org/10.1103/PhysRevLett.128.191602}{\emph{Phys. Rev. Lett.}
  {\bfseries 128} (2022) 191602}
  [\href{https://arxiv.org/abs/2109.09914}{{\ttfamily 2109.09914}}].

\bibitem{Banerjee:2022dgv}
A.~Banerjee, T.~Kibe, N.~Mittal, A.~Mukhopadhyay and P.~Roy, \emph{{Erasure
  tolerant quantum memory and the quantum null energy condition in holographic
  systems}}, \href{https://doi.org/10.1103/PhysRevLett.129.191601}{\emph{Phys.
  Rev. Lett.} {\bfseries 129} (2022) 19}
  [\href{https://arxiv.org/abs/2202.00022}{{\ttfamily 2202.00022}}].

\bibitem{Bzowski:2018pcy}
A.~Bzowski and M.~Guica, \emph{{The holographic interpretation of $J \bar
  T$-deformed CFTs}},
  \href{https://doi.org/10.1007/JHEP01(2019)198}{\emph{JHEP} {\bfseries 01}
  (2019) 198} [\href{https://arxiv.org/abs/1803.09753}{{\ttfamily
  1803.09753}}].

\bibitem{Papadimitriou:2007sj}
I.~Papadimitriou, \emph{{Multi-Trace Deformations in AdS/CFT: Exploring the
  Vacuum Structure of the Deformed CFT}},
  \href{https://doi.org/10.1088/1126-6708/2007/05/075}{\emph{JHEP} {\bfseries
  05} (2007) 075} [\href{https://arxiv.org/abs/hep-th/0703152}{{\ttfamily
  hep-th/0703152}}].

\bibitem{Witten:2001ua}
E.~Witten, \emph{{Multitrace operators, boundary conditions, and AdS / CFT
  correspondence}},  \href{https://arxiv.org/abs/hep-th/0112258}{{\ttfamily
  hep-th/0112258}}.

\bibitem{Balasubramanian:1999re}
V.~Balasubramanian and P.~Kraus, \emph{{A Stress tensor for Anti-de Sitter
  gravity}}, \href{https://doi.org/10.1007/s002200050764}{\emph{Commun. Math.
  Phys.} {\bfseries 208} (1999) 413}
  [\href{https://arxiv.org/abs/hep-th/9902121}{{\ttfamily hep-th/9902121}}].

\bibitem{Polchinski:2010hw}
J.~Polchinski, \emph{{Introduction to Gauge/Gravity Duality}},  in
  \emph{{Theoretical Advanced Study Institute in Elementary Particle Physics}:
  {String theory and its Applications: From meV to the Planck Scale}},
  pp.~3--46, 10, 2010, \href{https://doi.org/10.1142/9789814350525_0001}{DOI}
  [\href{https://arxiv.org/abs/1010.6134}{{\ttfamily 1010.6134}}].

\bibitem{Ecker:2019ocp}
C.~Ecker, D.~Grumiller, W.~van~der Schee, M.M.~Sheikh-Jabbari and P.~Stanzer,
  \emph{{Quantum Null Energy Condition and its (non)saturation in 2d CFTs}},
  \href{https://doi.org/10.21468/SciPostPhys.6.3.036}{\emph{SciPost Phys.}
  {\bfseries 6} (2019) 036} [\href{https://arxiv.org/abs/1901.04499}{{\ttfamily
  1901.04499}}].

\bibitem{Wall:2017blw}
A.C.~Wall, \emph{{Lower Bound on the Energy Density in Classical and Quantum
  Field Theories}},
  \href{https://doi.org/10.1103/PhysRevLett.118.151601}{\emph{Phys. Rev. Lett.}
  {\bfseries 118} (2017) 151601}
  [\href{https://arxiv.org/abs/1701.03196}{{\ttfamily 1701.03196}}].

\end{thebibliography}\endgroup
\bibliographystyle{JHEP}

\end{document}